\documentclass{ethpaper}
\usepackage{graphicx}
\usepackage{latexsym,rotating,subfigure}
\usepackage{subfigure}
\usepackage{wrapfig,epsfig}

\begin{document}

\begin{titlepage}

   \ethnote{}
\title{Crystals for High-Energy Physics calorimeters in extreme environments}

\begin{Authlist}
F.~Nessi-Tedaldi
\Instfoot{eth}{Swiss Federal Institute of Technology (ETH),
CH-8093 Z\"urich, Switzerland}
  \end{Authlist}
\maketitle
\abstract{Scintillating crystals are used for calorimetry in several
  high-energy physics experiments. For many of them, performance has
  to be ensured in very difficult operating conditions, like a high
  radiation environment and large particle fluxes, which place
  constraints on response and readout time.  An overview is presented
  of the knowledge reached up to date, and of the newest achievements
  in the field, with particular attention given to the performance of
  Lead Tungstate crystals exposed to large particle fluxes.}
\vspace{7cm}
\conference{Presented at the 9th ICATPP Conference on\\
Astroparticle, Particle, Space Physics,
 Detectors and Medical Physics Applications\\ Como, Italy,
October 17th to 21st, 2005}

\end{titlepage}

\setcounter{page}{2}

\section{Introduction}

This report addresses the performance of scintillating crystals used
for high-energy physics calorimetry, when operation implies high
radiation levels and intense particle fluxes.

The effect of high levels of ionising radiation on crystals has been
studied in depth and reported upon by many authors, as crystals were
used e.g. in $e^+e^-$ collider experiments, and their growth
parameters were optimised for best performance in such environments.
They are briefly summarised herein.  Hadron collider detectors today
share the same concern, but add to it the need to ensure adequate
performance when crystals are exposed to large particle fluxes. Such
running conditions are namely expected in several experiments under
construction or designed. Some new results are thus presented here,
together with a fresh look at existing, older ones, to provide, as far
as possible, a complete picture.

\section{Performance under high ionising radiation levels}

Ionising radiation is known to produce absorption bands through
formation of colour centres, which reduce the Light Transmission (LT)
and thus the Light Output (LO), due to oxygen contamination in alkali
halides like $\mathrm{BaF}_2$ and CsI, and to oxygen vacancies and
impurities in oxides like BGO and
$\mathrm{PbWO}_4$~\cite{r-ZHU1}. Phosphorescence or afterglow appear
sometimes\cite{r-ETH1}, which increase the noise levels in the
detected light signal, possibly worsening the energy resolution (in a
negligible way for $\mathrm{PbWO}_4$ in LHC
experiments\cite{r-ZHU2}), while the scintillation mechanism is
generally not damaged.
Recovery of damage at room temperature can occur depending on crystal
type and growth parameters, giving rise to a dose-rate dependence of
damage equilibrium levels\cite{r-ZHU1,r-ETH2} and to a recovery speed
dependent on the depth of traps.  That ionising radiation only affects
LT, means the damage can be monitored through light injection and
corrected for, as it is done in the CMS Electromagnetic Calorimeter
(ECAL)\cite{r-ZHANG}.

\section{Performance in large particle fluxes}

The way hadron fluxes affect crystals has become a crucial question
while detectors making use of this calorimetry technique are being
constructed.  In particular, it had to be ascertained whether such
fluxes cause a specific, possibly cumulative damage, and if so, what
its quantitative importance is, whether it only affects LT or also the
scintillation mechanism. Extensive studies have been recently
performed on $\mathrm{PbWO}_4$ at IHEP Protvino\cite{r-Batarin} and,
for the CMS ECAL, at CERN and ETH-Z\"urich\cite{r-ETH3}. Their main 
results are quoted and discussed herein.

Crystal tests at Protvino were using $e^-$ and $\pi$ beams and
$\gamma$ sources up to a few krad at 1 to 60 rad/h at one end, and a
very intense mixed beam of charged hadrons, neutrons and $\gamma$ up
to 3 Mrad at 1 krad/h and 100 krad/h equivalent fluxes at the other
end. In individual $e^-$, $\pi$ and $\gamma$ irradiations, the signal
loss behaviour is found to be qualitatively similar between electrons
and pions, and the damage appears to reach equilibrium at a dose-rate
dependent level. Furthermore, no indication of damage to the
scintillation mechanism from $\pi$ irradiation is
found\cite{r-Batarin3}. A concern remains however, that an
additional, specific, possibly cumulative damage from hadrons cannot
be excluded and could appear when a high total integrated dose is reached.
This concern is partially confirmed by irradiations in the very
intense, mixed beam. Under the constant flux used, the damage appears
in fact to be steadily increasing with accumulated dose. This is
unlike pure ionising radiation damage, which reaches equilibrium at a
level depending on dose rate, not beyond what saturation of all colour
centres can yield. Therefore, this constitutes an indication for a
cumulative, hadron-specific damage.

For CMS, hadron fluences have been calculated\cite{r-ETH4} for
$5\times 10^5\;\; \mathrm{pb}^{-1}$ (10 y running at LHC), yielding in
the ECAL barrel (end caps) $\sim 10^{12}\; (\sim 10^{14})\; $charged
hadrons/$\mathrm{cm}^2$ .  A hadron-specific damage could arise from
the production, above a $\sim 20$ MeV threshold, of heavy fragments
(``stars''), with up to 10 $\mu$m range and energies up to $\sim$100
MeV, causing a displacement of lattice atoms and energy losses along
their path up to 50000 times the one of minimum-ionising
particles. The damage caused by these processes is likely different
from the one of ionising radiation, thus possibly cumulative. The
primarily investigated quantity was the damage to Light Transmission
measured longitudinally through the length (L) of
the crystal and quantified as the induced absorption coefficient at
peak-of-emission wavelength,
$\mu_{IND}\mathrm{(420\; nm)}=\frac{1}{L}\ln\frac{LT_{INIT}}{LT_{END}}$,
with $LT_{INIT}$ and $LT_{END}$ the longitudinal Light Transmission at 420 nm
before and after irradiation.
The Perkin Elmer $\Lambda_{900}$ spectrophotometer used, allows in fact
to measure LT very accurately, to better than 1\%.
Transmission is furthermore related to LO changes, provided scintillation
is not affected.

Eight CMS production crystals of consistent quality were irradiated at
the IRRAD1 facility of the CERN PS accelerator T7 beam
line\cite{r-GLA} in a 20 GeV/c proton flux of
\begin{figure}[h]
\begin{center}\footnotesize
  \begin{tabular}[h]{cc}
  \subfigure[For proton-induced damage.\label{f-LTp}]
    {\mbox{\includegraphics[width=65mm,clip=true]{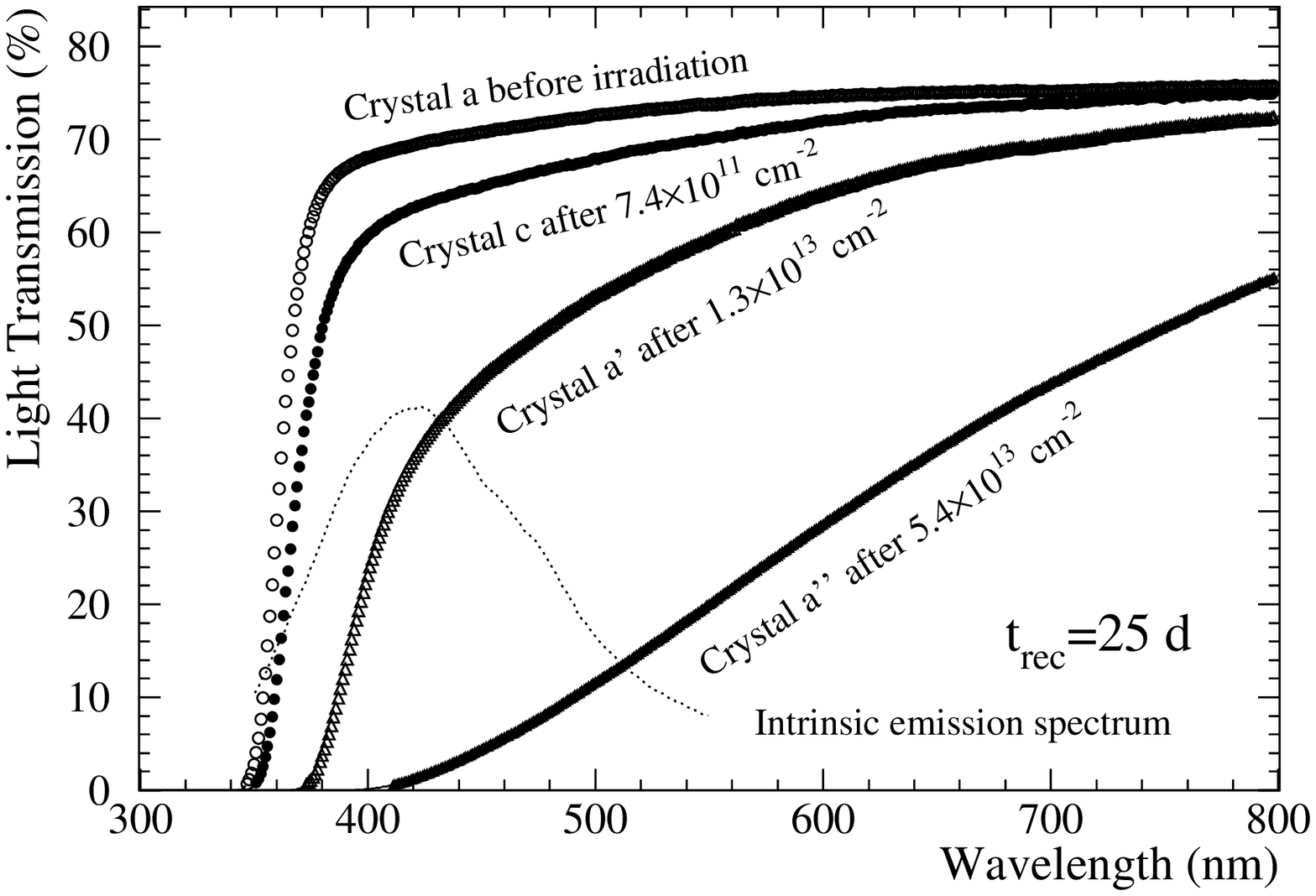}}}&
  \subfigure[For $\gamma$-induced damage. The top thin line shows LT
             prior to irradiation.\label{f-LTg}]
    {\mbox{\includegraphics[width=40mm,viewport= 28 17 280 280, clip=true]{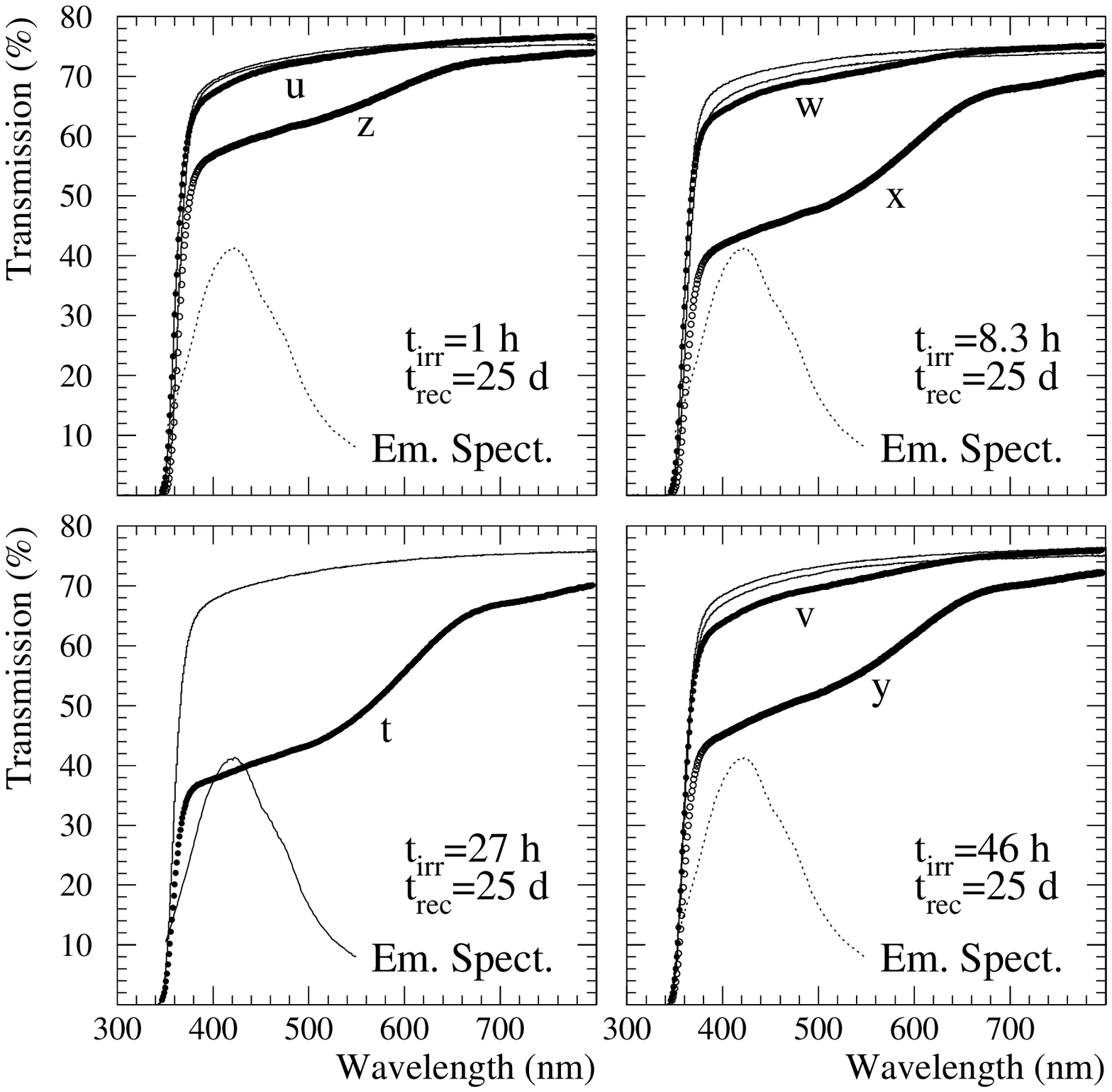}}}
  \end{tabular}
\end{center}
\caption{Longitudinal Light Transmission curves for crystals with
various degrees of radiation damage.\label{f-LT}}
\end{figure}
$10^{12}\;\mathrm{p/cm}^{2}/\mathrm{h}$ (crystals {\em a'', b, c, d, e, h}) or of
$10^{13}\;\mathrm{p/cm}^{2}/\mathrm{h}$ (crystals {\em E, F', G})
\footnote{Prime (respectively '') indicates a second (or third)
irradiation of the same crystal.}.
To disentangle the contribution to damage from the associated ionising
dose, complementary $^{60}\mathrm{Co}\; \gamma$-irradiations were
performed at a dose rate of 1 kGy/h on seven further crystals
({\em t, u, v, w, x, y, z}) at the ENEA Casaccia Calliope plant\cite{r-ENEA}.
In fact, a flux of $10^{12}\;\mathrm{p/cm}^{2}/\mathrm{h}$ has an
associated ionising dose rate in $\mathrm{PbWO}_4$ of 1 kGy/h.
The LT data in Fig.~\ref{f-LTp} show a smooth worsening of LT with
increasing proton fluence over the entire range of wavelengths, and a
clear shift of the Transmission band-edge. In $\gamma$-irradiated
crystals (Fig.~\ref{f-LTg}, where also the emission
spectrum\cite{r-emi} is indicated), the band-edge does not shift at
all, even after the highest cumulated dose reached: just the usual
absorption band appears around 420 nm. These data thus give prominence
to the qualitatively different fundamental nature of proton-induced
and $\gamma$-induced damage.

\begin{figure}[h]
\begin{center}\footnotesize
\begin{tabular}[h]{ccc}
\subfigure[Induced absorption $\mu_{IND}\mathrm{(420\; nm)}$ as a
function of cumulated proton fluence.\label{f-muvsflu}]
{\mbox{\includegraphics[width=6cm]{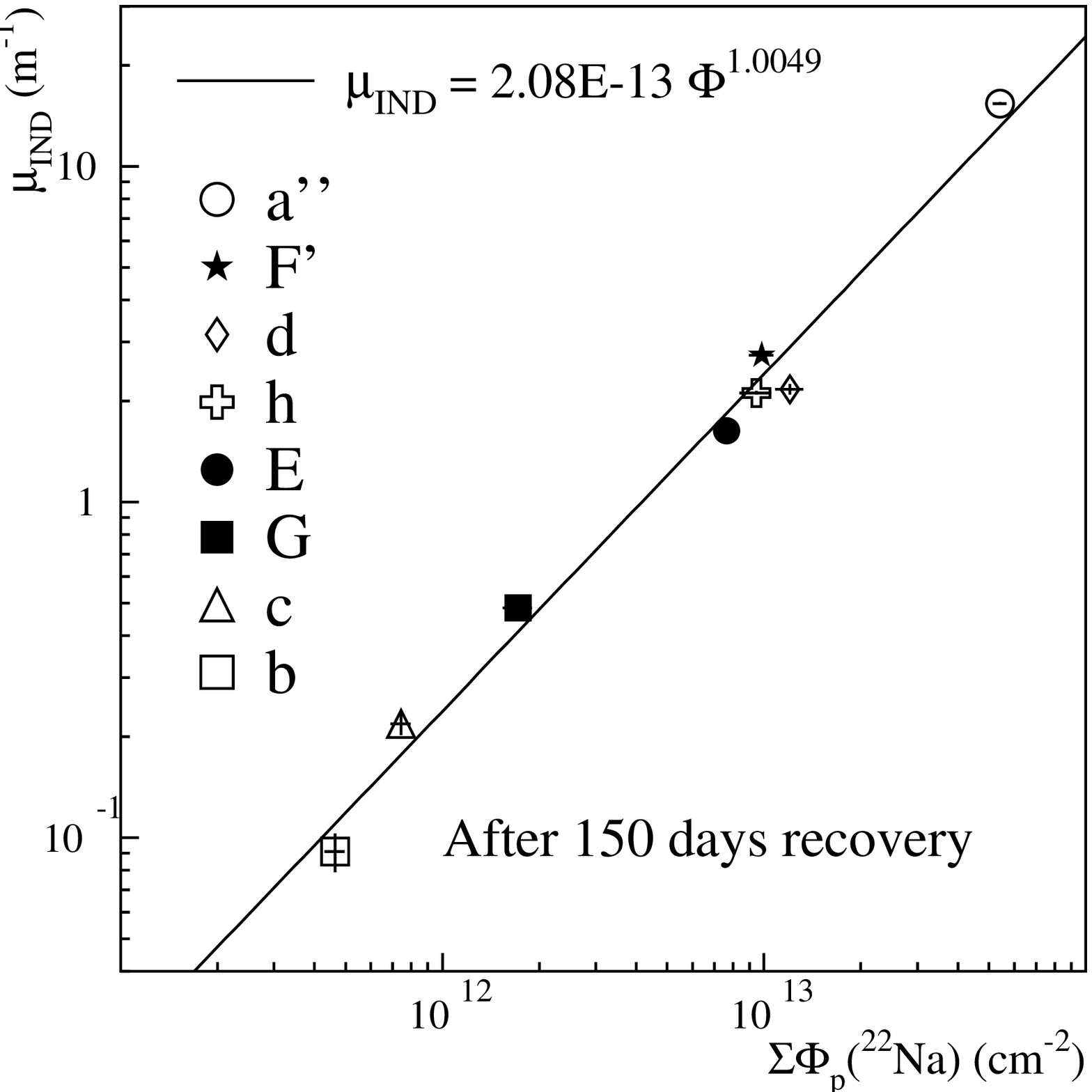}}} &
{\mbox{\includegraphics[height=54mm,viewport= 4 14 54 260, clip=true]{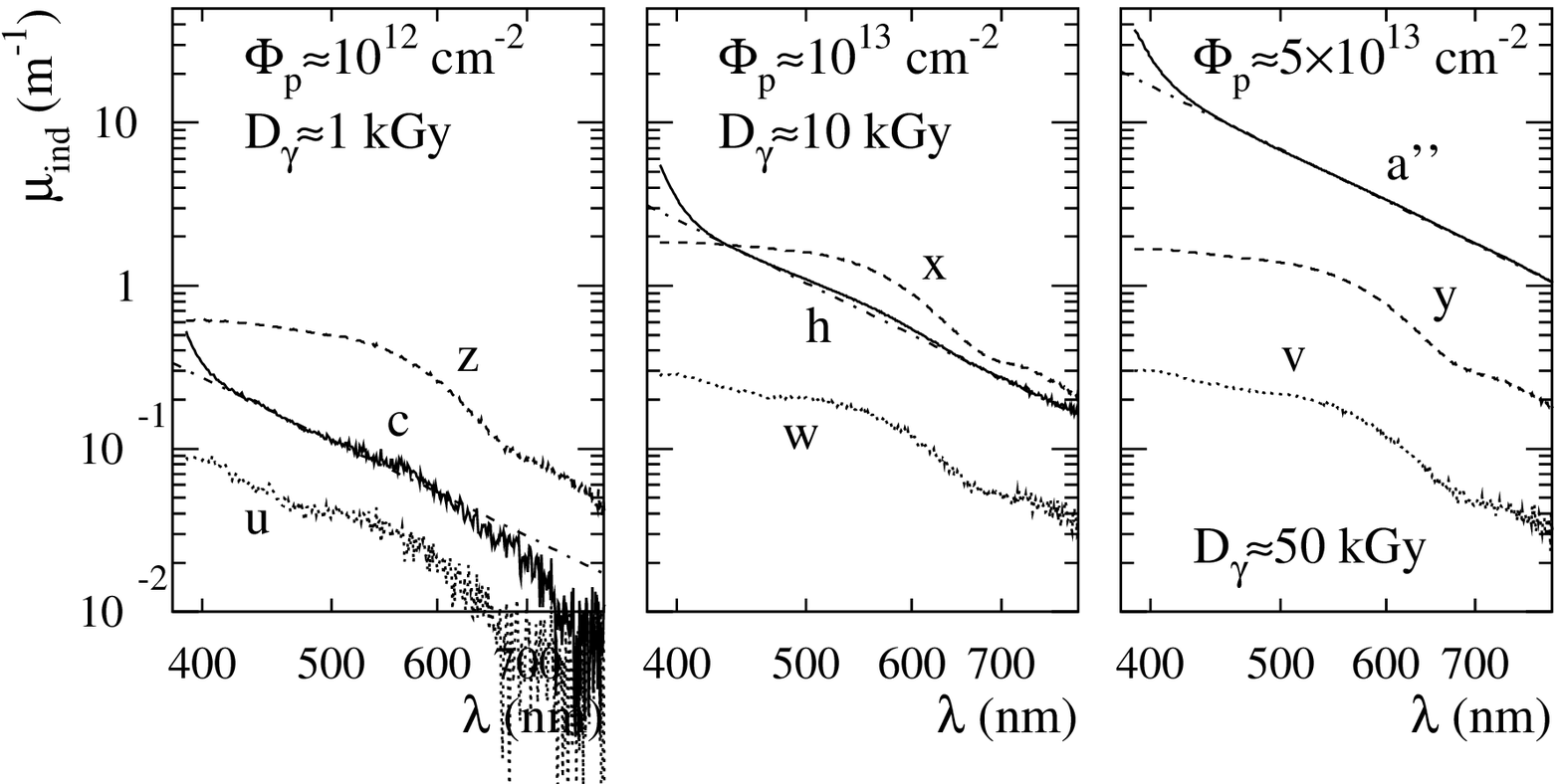}}} &
\subfigure[$\mu_{IND}\mathrm{(420\; nm)}$ as a function of $\lambda$\label{f-muvsl}]
{\mbox{\includegraphics[height=54mm,viewport= 360 14 513 260, clip=true]{fig17NC.eps}}}
\end{tabular}
\end{center}
\caption{Behaviour of irradiation damage for crystals irradiated
with protons ({\em a'', b, c, d, E, F', G, h}) and with
$\gamma$ ({\em v, y}).\label{fig1}}
\end{figure}

The correlation in Fig.~\ref{f-muvsflu}, between
$\mu_{IND}\mathrm{(420\; nm)}$ and fluence, is consistent with a linear
behaviour over two orders of magnitude, showing that proton-induced
damage in $\mathrm{PbWO}_4$ is predominantly cumulative, unlike
$\gamma$-induced damage, which reaches
equilibrium\cite{r-ZHU1,r-ETH2}.  Figure~\ref{f-muvsl} shows
$\mu_{IND}\mathrm{(420\; nm)}$ plotted versus light wavelength for the
proton-irradiated crystal {\em a''} and for the two $\gamma$ -
irradiated crystals {\em v} and {\em y}.  The dot-dashed line shows
$\lambda^{-4}$ fitted to the data of the proton-damaged crystal {\em
  a''}. 
The good agreement is an indication of Rayleigh scattering
from small centres of severe damage. This is consistent with an origin
of damage due to the high energy deposition of heavily ionising
fragment along their path, that changes locally the crystal structure.
Taking into account the difference in composition and energy spectra
between 20 GeV/c protons and CMS, simulations indicate that the test
\begin{figure}[ht]
\begin{center}\footnotesize
\mbox{\includegraphics[width=8cm]{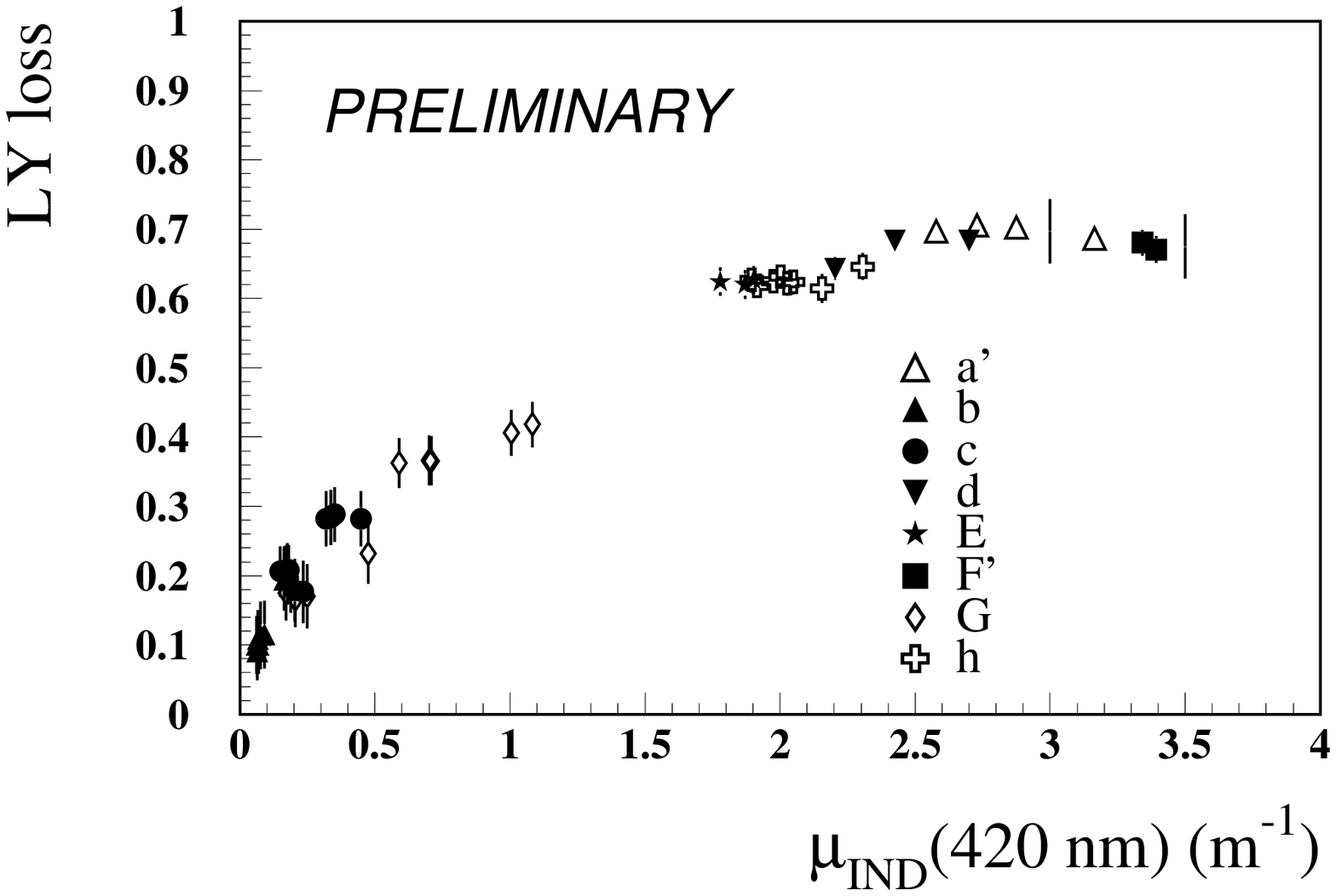}}\\
\mbox{\includegraphics[width=8cm]{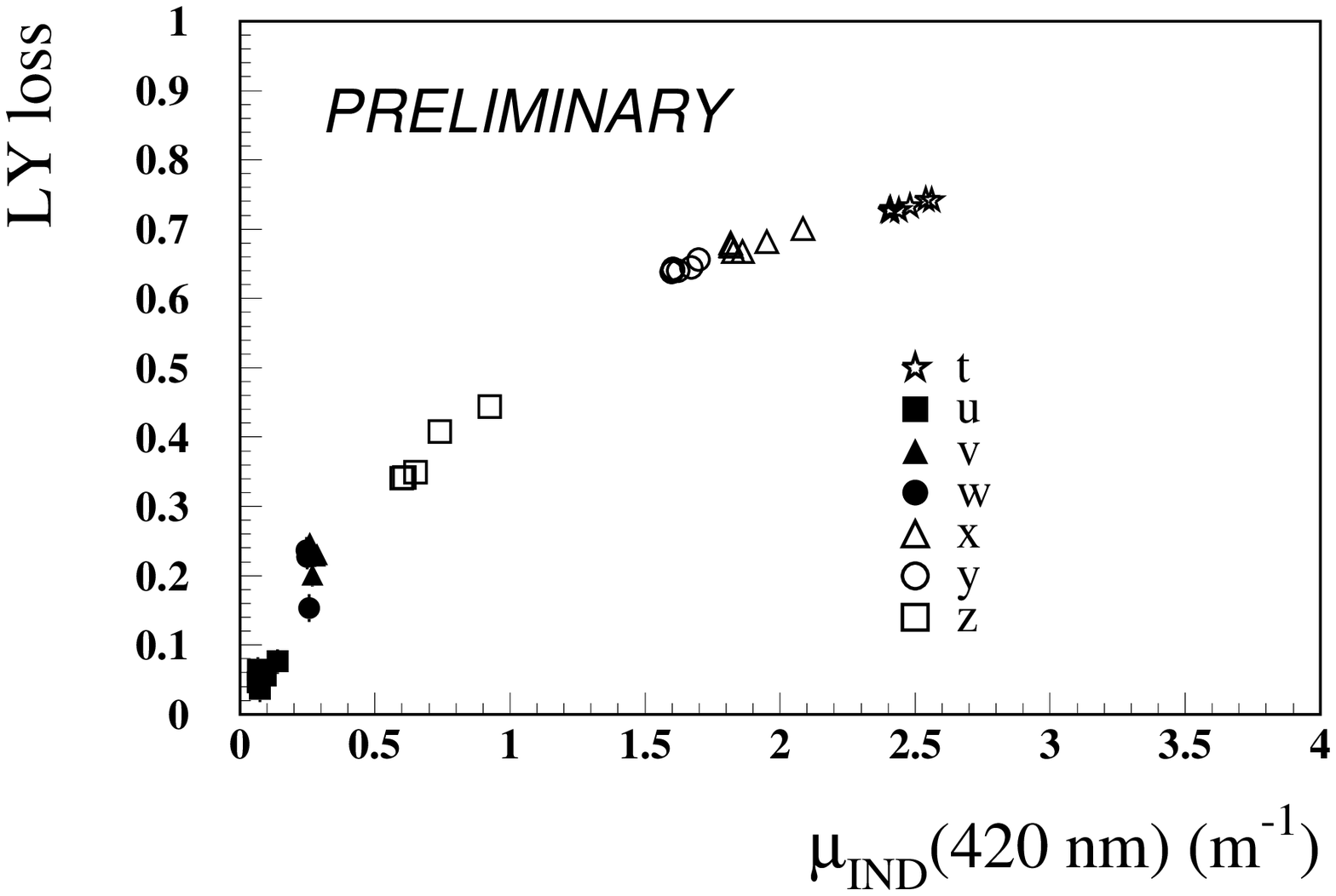}}
\end{center}
\caption{Correlation between $\mu_{IND}\mathrm{(420\; nm)}$ and Light Output
loss for proton-induced (top) and $\gamma$-induced (bottom) damage.
 \label{f-LYcorrel}}
\end{figure}
results cover the CMS running conditions up to $\sim 2.6$. An
experimental confirmation is expected in the future from a
pion-irradiation of $\mathrm{PbWO}_4$, closely approximating the CMS
particle spectrum and energies.

The evolution of Light Output was also monitored on the same set of
irradiated crystals, using cosmic muons, traversing the crystals
transversely and thus leaving approximately 30 MeV of energy deposit,
to excite scintillation.  The correlation\cite{r-ETH5} between
$\mu_{IND}\mathrm{(420\; nm)}$ and Light Output loss is shown in the top
part of Fig.~\ref{f-LYcorrel} for all proton-irradiated crystals, and
at the bottom for all $\gamma$-irradiated ones. The vertical bars
indicate the systematic scale uncertainty affecting the data for {\em
  a'} and {\em F'}. For both, proton-irradiated and
$\gamma$-irradiated crystals, the measured Light Output loss
correlates well with $\mu_{IND}\mathrm{(420\; nm)}$.
Furthermore, within the precision of the measurements, no difference
can be observed in this correlation between the two sets of crystals
and thus no hadron-specific alteration of the scintillation properties
can be claimed.
 
\begin{figure}
\begin{center}
\mbox{\includegraphics[width=85mm]{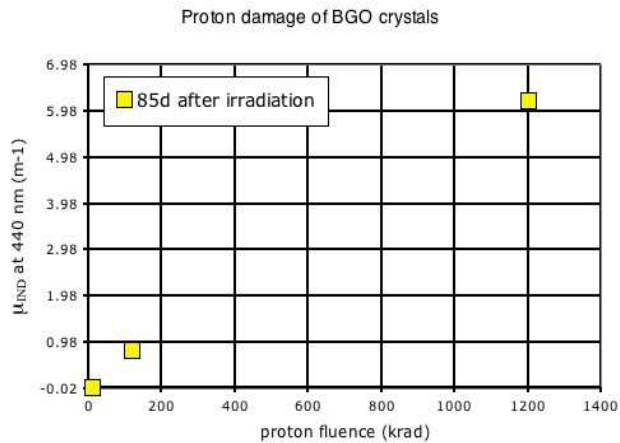}}
\end{center}
\caption{Correlation between $\mu_{IND}\mathrm{(440\; nm)}$ and proton
fluence in BGO extracted from published data (see text).\label{f-BGO}}
\end{figure}

Proton and $\gamma$ data are also compared in a study performed on
BGO\cite{r-KobaBGO}. The changes in band-edge are similar to what is
seen in $\mathrm{PbWO}_4$, and long enough after irradiation, when the
ionising-radiation damage contribution has recovered, one can extract
a remaining proton-induced damage that behaves linearly with fluence,
as visible in Fig.~\ref{f-BGO}.  The same exercise is not possible on
CsI data from the same authors\cite{r-KobaCsI} because the damage
caused by ionising radiation gives a contribution which is too
important to allow disentangling the proton-specific one.

In conclusion, one can say that for all crystals commonly used in
calorimetry, beyond the well-studied damage from ionising radiation,
the understanding of additional contributions to the damage, when
crystals experience a substantial hadron flux, has become important
since experiments are being built having to cope with such running
conditions. A hadron-specific, cumulative contribution, likely due to
the intense local energy deposition from heavy fragments, has been
observed in $\mathrm{PbWO}_4$ and BGO. Over the explored flux and
fluence ranges and within the accuracy of the measurements, this
contribution is observed to only affect Light Transmission, and thus
can be monitored through light injection. 
Additional studies are expected to consolidate the present
understanding of hadron damage.

\end{document}